\documentstyle[12pt]{article}

\catcode`\@=11 \@addtoreset{equation}{section}\catcode`\@=12
\newcommand{\be}{\begin{equation}}
\newcommand{\ee}{\end{equation}}
\newcommand{\ba}{\begin{eqnarray}}
\newcommand{\ea}{\end{eqnarray}}

\begin{document}
\thispagestyle{empty}

\begin{center}
               RUSSIAN GRAVITATIONAL SOCIETY\\
               CENTER FOR SURFACE AND VACUUM RESEARCH\\
               DEPARTMENT OF FUNDAMENTAL INTERACTIONS AND METROLOGY\\
\end{center}
\vskip 4ex
\begin{flushright}                              RGS-CSVR-002/95\\
                                                gr-qc/9503223

 \end{flushright}
\vskip 15mm

\begin{center}
{\large\bf
Multidimensional Classical and Quantum Cosmology
with Perfect Fluid}

\vskip 5mm
{\bf
V. D. Ivashchuk and V. N. Melnikov }\\
\vskip 5mm
     {\em Center for Surface and Vacuum Research,\\
     8 Kravchenko str., Moscow, 117331, Russia}\\
     e-mail: mel@cvsi.rc.ac.ru \\
\end{center}
\vskip 10mm

ABSTRACT

A cosmological model describing the evolution of $n$ Ricci-flat
spaces $(n>1)$ in the presence of  $1$-component perfect-fluid
and  minimally coupled scalar field is considered. When the
pressures in all spaces are proportional to the density: $p_{i} =
(1- h_{i}^{}) \rho$, $h_{i} = const$, the Einstein and
Wheeler-DeWitt equations are integrated for a large variety of
parameters $h_{i}$. Classical and quantum wormhole solutions are
obtained for negative density $\rho < 0$. Some special classes of
solutions, e.g. solutions with spontaneous and dynamical
compactification, exponential and power-law inflations, are
singled out. For $\rho > 0$  a third quantized cosmological model
is considered and the Planckian spectrum of "created universes" is
obtained.

\vskip 10mm


\vskip 30mm

\centerline{Moscow 1995}
\pagebreak

\setcounter{page}{1}

\pagebreak

\section{Introduction}
\setcounter{equation}{0}

Last years the  of Kaluza-Klein ideas \cite{Kal,Kl}
(see also \cite{DeS}-\cite{WeP}) and  superstring theory
\cite{GrSW} greatly stimulated the interest in multidimensional cosmology
(see, for example, \cite{BelKh}-\cite{Ang} and references
therein).

Classical and quantum multidimensional cosmological models were
investigated in our papers  starting from  \cite{IM1}-\cite{IMZ}
(and for spherical symmetry case from \cite{BrI,FIM2}).
Some windows to observational effects of extra
dimensions were found and analyzed such as possible variations of the
effective gravitational constant, its relations with other cosmological
parameters \cite{IM1,BIM,FIM1}.

But the treatment of classical models may be only the necessary
first step in analyzing the properties of the "early universe" and
last stages of the gravitational collapse in a multidimensional
approach. [The content of this paragraph is a personal opinion of
the second author (V.N.M.).] In quantum multidimensional cosmology
we hope to find answers to such questions as the singular state,
the "creation of the Universe", the nature and value of the
cosmological constant, some ideas about possible "seeds" of the
observable structure of the Universe, stability of fundamental
constants etc. In the third quantization scheme the problems of
topological changes may be treated thoroughly. It should be noted
also that the multidimensional schemes may be also used in
multicomponent inflationary scenarios \cite{Lin}- \cite{StPol}
(see for example \cite{GasVen1}- \cite{Ang}).

In this paper we consider a cosmological model describing the
evolution of $n$ Ricci-flat spaces $(n>1)$
with  $1$-component perfect-fluid  and  minimally coupled scalar field as
a matter source.  The pressures in all space are proportional to the
density of energy and the coefficients of proportionality satisfy
a certain inequality (see (2.8) and (2.13)).
This model is investigated  in classical, quantum
and third quantized cases and  corresponding exact solutions are found.
Some particular
models of such type were considered previously by many authors
(see, for example, \cite{Sah,GRT,BL1,BL2,BL3,IM2,DP}). In
\cite{IM5} some classes of exact solutions to Einstein and
Wheeler-DeWitt equations with multicomponent perfect fluid
and a chain of Ricci-flat internal spaces
were obtained. In \cite{GIM} new families
of classical solutions for the model \cite{IM5}
(including those of Toda-like type) were considered.

The paper is organized as following. In Sec. 2
the general description of the model is performed.
In Sec. 3 we integrate the Einstein equations for the model
and analyze a class of exceptional (inflationary) solutions.
(The solutions with $n =2$ were considered recently in
\cite{GasVen2,Ang}).
The isotropization-like and Kasner-like asymptotical behaviours
of the solutions are analized.
In subsection 3.5 some special cases such as isotropic
(when the pressures in all spaces
are equal) and curvature-like cases are investigated.
In the last case there exist solutions with so-called
spontaneous and dynamical compactifications.
The instanton solutions
(classical wormholes) with imaginary scalar field and
negative energy density are also obtained.
(The interest to wormhole solutions was stimulated
greatly after the papers \cite{GidS1,My,Haw}).
We note that the exact solutions for
1-component perfect fluid (without scalar field) and
a chain of Ricci-flat spaces, were obtained for
the first time in \cite{IM2} (see also \cite{IM5}).
The cosmological constant case was considered previously in
\cite{IM4,BIMZ}.

In Sec. 4  we consider our model at the quantum level
(for  pioneering papers see \cite{ADM}-\cite{StM}).
Here we quantize scale  factors and a scalar field but treat
the perfect fluid as a classical object. Such approach
is quite consistent at least in  certain  special
situations  such as $\Lambda$-term  \cite{IM4,BIMZ}
and curvature \cite{Zh1}-\cite{Zh4} cases.

In Sec. 4 the  Wheeler-DeWitt equation for the model
is solved and quantum wormhole solutions are obtained.
We recall that the notion of quantum wormholes
was introduced by Hawking and Page as a quantum
extension of the classical wormhole paradigma (see also
\cite{CGar}-\cite{Mar} and  \cite{Zh2,Zh3,IM4,BIMZ}).
They proposed to regard
quantum wormholes  as solutions of the Wheeler-DeWitt (WDW) equation
with the following boundary conditions:
(i) the wave function is
exponentially damped for large "spatial geometry";
(ii) the wave
function is regular when the spatial geometry degenerates.
The first condition expresses the fact that a space-time should be
Euclidean at the spatial infinity. The second condition should reflect the
fact that the space-time is nonsingular when a spatial geometry
degenerates.  (For example, the wave function should not oscillate an
infinite number of times).  Presented in this paper multidimensional
quantum wormhole solutions may be considered as a natural extension of the
corresponding solutions in \cite{Zh2,Zh3}  and  \cite{IM4,BIMZ} for
curvature and $\Lambda$-term cases correspondingly.

In Sec. 5  a third quantized cosmology is investigated along a
line as it was done in  \cite{Zh1} and  \cite{Zh5}  for curvature
and cosmological constant cases correspondingly. Here we are lead
to the theory of massless conformally coupled scalar field  in
conformally flat generalized Milne universe \cite{Zh5}. In- and
out-vacuums are defined and planckian spectrum for the "created
out-universes" (from in-vacuum) is obtained using standard
relations \cite{BirD,GrMM}. The temperature is shown to depend
upon the equation of state. It should be noted that recently the
interest to the third quantized models was stimulated by papers
\cite{Rub,GidS2} (see also \cite{Kir}-\cite{Hor} and references
therein).

\section{The model}
\setcounter{equation}{0}

We consider a cosmological model describing
the evolution of $n$ Ricci-flat spaces in the presence of the
$1$-component perfect-fluid matter \cite{IM5} and a homogeneous massless
minimally coupled scalar field.  The metric of the model
\begin{equation}
g=-exp[2{\gamma}(t)]dt \otimes dt +
\sum_{i=1}^{n} exp[2{x^{i}}(t)]g^{(i)},
\end{equation}
is defined on the  manifold
\begin{equation}
M = R \times M_{1} \times \ldots \times M_{n},
\end{equation}
where the manifold $M_{i}$ with the metric $g^{(i)}$ is a
Ricci-flat space of dimension $N_{i}$, $i = 1, \ldots ,n $; $n \geq 2$.
We take the field equations in the following form:
\ba 
&&R^{M}_{N}-\frac{1}{2}\delta^{M}_{N}R = \kappa^{2}T^{M}_{N}, \\
&& {\Box} {\varphi} = 0,
\ea
where $\kappa^{2}$ is the gravitational constant,
$ \varphi = {\varphi}(t)$ is scalar field,  ${\Box}$ is the d'Alembert
operator for the metric (2.1) and
the energy-momentum tensor is adopted in the following form
\begin{eqnarray} 
&& T^{M}_{N} =  T^{M (pf)}_{N}  + T^{M (\phi)}_{N} , \\
&&(T^{M (pf)}_{N})= diag(- \rho,
 p_{1} \delta^{m_{1}}_{k_{1}},
\ldots , p_{n} \delta^{m_{n}}_{k_{n}}), \\
&&T^{M(\phi)}_{N}  = \partial^M \varphi \partial_N \varphi
- \frac{1}{2} \delta^M_N (\partial \varphi)^2.
\end{eqnarray}
We put  pressures of the perfect fluid in all spaces to be
proportional to the density
\begin{equation} 
{p_{i}}(t) = (1- \frac{u_{i}}{N_{i}}) {\rho}(t),
\end{equation}
where $u_{i} = const$, $i =1, \ldots ,n$.

We impose also the following restriction on the vector $u = (u_i) \in
R^{n}$
\be 
<u,u>_{*} < 0.
\ee
Here bilinear form $<.,.>_{*}: R^{n}
\times R^{n} \rightarrow R$ is defined by the relation
\be 
<u,v>_{*} = G^{ij} u_i v_j,
\ee
$u,v \in R^{n}$, where
\begin{equation}  
G^{ij} = \frac{\delta^{ij}}{N_{i}} + \frac{1}{2-D}
\end{equation} are components
of the matrix inverse to the matrix of the minisuperspace metric
\cite{IM2,IMZ}
\begin{equation} 
G_{ij} = N_{i} \delta_{ij} - N_{i}N_{j}.
\end{equation}
In (2.11) $D = 1 + \sum_{i=1}^{n} N_i$ is the
dimension of the manifold $M$ (2.2).

Remark 1. This restriction (2.9) reads
\begin{equation}
<u,u>_{*}  \equiv G^{ij}u_{i}u_{j}=
\sum_{i=1}^{n} \frac{(u_{i})^{2}}{N_{i}}
+ \frac{1}{2-D}(\sum_{i=1}^{n} u_{i})^{2} < 0.
\end{equation}
We note that in notations of \cite{IM2} $<u,u>_{*} = {\acute{\Delta}}(h)/
(2-D)$.)

\section{Classical solutions}

The Einstein equations (2.3) imply
$\bigtriangledown_{M} T^{M}_{N}=0$ and due to (2.4)
$\bigtriangledown_{M}T^{M (pf)}_{N} = 0$ or equivalently
\begin{equation} 
\dot{\rho}+ \sum_{i=1}^{n} N_{i} \dot{x}^{i}(\rho + p_{i})=0.
\end{equation}
{}From (2.8), (3.1) we get
\begin{equation} 
{\kappa^{2} \rho}(t)=A \exp[-2N_{i}{x^{i}}(t) + u_i {x^i}(t)],
\end{equation}
where $A=const$. We put $A \neq 0$ (the case $A = 0$ was considered
thoroughly in \cite{BZ3}).

It is  not difficult to verify that the field equations (2.3), (2.4)
for the cosmological metric (2.1) in the harmonic time gauge
\begin{equation} 
\gamma_{0} \equiv \sum_{i=1}^{n} N_{i} x^{i}
\end{equation}
with the energy-momentum tensor from (2.5)-(2.7) and
the relations (2.8), (3.2) imposed are equivalent
to the Lagrange equations for the Lagrangian
\begin{equation} 
L = \frac{1}{2} (G_{ij} \dot{x}^{i} \dot{x}^{j} +
\kappa^{2} \dot{\varphi}^2)  - \kappa^{2} A \exp(u_k x^k)
\end{equation}
with the energy constraint
\begin{equation} 
E = \frac{1}{2} ( G_{ij} \dot{x}^{i} \dot{x}^{j} +
\kappa^{2} \dot{\varphi}^2)  + \kappa^{2} A \exp(u_k x^k) = 0
\end{equation}
(for $\varphi = 0$ see \cite{IM5,IM6}).

We recall \cite{IM2,IMZ} that the minisuperspace metric
\begin{equation} 
G = G_{ij}dx^{i}\otimes dx^{i}
\end{equation}
has the pseudo-Euclidean signature $(-,+, \ldots ,+)$, i.e. there
exists a linear transformation
\begin{equation}  
z^{a}= e^{a}_{i} x^{i},
\end{equation}
diagonalizing the minisuperspace metric (3.6), (2.12)
\begin{equation}  
G= \eta_{ab} dz^{a} \otimes dz^{b} =
- dz^{0} \otimes dz^{0} + \sum_{i=1}^{n-1} dz^{i} \otimes dz^{i},
\end{equation}
where
\begin{equation}  
(\eta_{ab})=(\eta^{ab}) \equiv diag(-1,+1, \ldots ,+1),
\end{equation}
$a,b = 0, \ldots ,n-1$.
{}From (3.7)-(3.8) we get
\begin{equation} 
\eta_{ab} e^{a}_{i} e^{b}_{j}= G_{ij}
\end{equation}
and as a consequence
\begin{equation} 
e_{a}^{i} G_{ij} e_{b}^{j}  = \eta_{ab}, \qquad
e_{a}^{i} = \eta_{ab} G^{ij} e^{b}_{j},
\end{equation}
where $(e_{a}^{i}) = (e^{a}_{i})^{-1}$.
As in \cite{IM5} we put
\be 
e^{0}_{i}= u_{i}/(2q) \Longrightarrow
z^0 = u_{i} x^i /(2q),
\ee
where here and below
\be 
2q \equiv \sqrt{- <u,u>_{*}}.
\ee
It may be done, since $<.,.>_{*}$ is a bilinear symmetrical
2-form of the signature $(-,+, \ldots ,+)$ and $<u,u>_{*} < 0$
\cite{IM5}.  An example of diagonalization (3.7) satisfying (3.12)
was considered in \cite{IM2,IMZ}.
{}From (3.11) and (3.12) we get
\be 
e_{0}^{i}= - G^{ij} e^{0}_{j} = - u^{i}/(2q), \qquad
u^{i} \equiv  G^{ij} u_{j}.
\ee

We also denote
\be 
z^n = \kappa \varphi.
\ee

The Lagrangian (3.4) in $z$-variables (3.7), (3.15) (with the relation
(3.12) imposed) may be rewritten as
\begin{equation}  
L = \frac{1}{2} \eta_{AB} \dot{z}^{A} \dot{z}^{B}
- \kappa^{2} A \exp(2qz^{0}),
\end{equation}
where $A,B = 0, \ldots ,n$. The energy constraint (3.5) reads
\begin{equation}  
E = \frac{1}{2} \eta_{AB} \dot{z}^{A} \dot{z}^{B}
+ \kappa^{2} A \exp(2q z^{0}) = 0.
\ee

The Lagrange equations for the Lagrangian (3.16)
\begin{eqnarray} 
&&-\ddot{z}^{0} + 2q A \exp(2q z^{0}) = 0, \\
&&\ddot{z}^{B} = 0, \qquad B  = 1, \ldots, n,
\end{eqnarray}
with the energy constraint  (3.17) can be easily solved.
{}From (3.19) we have
\begin{equation}  
z^{B} = p^{B} t + q^{B},
\end{equation}
where $p^{B}$ and $q^{B}$ are constants and $B= 1, \ldots , n$.
The first integral for  eq. (3.18) reads
\be  
- \frac{1}{2} (\dot z^0)^2 +  A \exp(2qz^0) + {\cal E} = 0.
\ee
Using (3.17), (3.20) and (3.21) we get
\begin{equation} 
{\cal E} =  \frac{1}{2} \sum_{B=1}^{n} (p^{B})^{2}.
\end{equation}
We obtain the following solution of eqs. (3.18), (3.21)
\begin{eqnarray}  
\exp(-2qz^{0})  = &(A/{\cal E}) \sinh^{2}(q \sqrt{2{\cal E}}(t-
t_{0})),  \qquad {\cal E} > 0, \ A > 0,  \\
&(A/{\cal |E|}) \sin^{2}(q \sqrt{2|{\cal E}|}(t- t_{0})), \qquad
{\cal E} < 0, \ A > 0,  \\
&2 q^2 A (t- t_{0})^2, \ \ \qquad {\cal E} = 0, \ A > 0,  \\
&(|A|/{\cal E}) \cosh^{2}(q \sqrt{2{\cal E}}(t- t_{0})),
\qquad {\cal E} > 0, \ A < 0,
\end{eqnarray}
Here $t_{0}$ is an arbitrary
constant.  For real  $z^B$ (or, equivalently, for the real metric and
the scalar field) we get from (3.22) ${\cal E} \geq 0$. The case ${\cal E}
< 0$   may take place when a pure imaginary scalar field is considered.

\subsection{Kasner-like parametrization for  non-exceptional
solutions with  real scalar field}

First, we consider the
real case with ${\cal E} > 0$. In this case the relations (3.23) and
(2.26) may be written in the following form
\begin{equation} 
\exp(-2qz^{0}) =
\frac{|A|}{{\cal E}} {f^{2}_{\delta}}(q \sqrt{2{\cal E}}(t- t_{0})),
\end{equation}
where $\delta \equiv  A / |A| = \pm 1$ and
\begin{eqnarray} 
{f_{\delta}}(x) \equiv \frac{1}{2} (e^{x} - \delta e^{-x})
& = &\sinh x, \qquad \delta = +1, \nonumber \\
& = &\cosh x, \qquad \delta = -1.
\end{eqnarray}
We introduce a new
time variable by the relation
\begin{eqnarray}  
\tau = && \frac{T}{\sqrt{\delta}} \ln \frac
{\exp[q \sqrt{2{\cal E}}(t - t_{0})] + \sqrt{\delta}}
{\exp[q \sqrt{2{\cal E}}(t - t_{0})] - \sqrt{\delta}} =   \\
&&T \ln \coth[\frac{1}{2} q \sqrt{2{\cal E}}(t - t_{0})], \qquad
\delta = +1,  \\
&&2T \arctan \exp[ - q \sqrt{2{\cal E}}(t - t_{0})], \qquad
\delta = -1,
\end{eqnarray}
where
\begin{equation} 
T = {T}(u,A) \equiv (2 q^2 |A|)^{- 1/2} = (\frac{1}{2}|A<u,u>_{*}|)^{-1/2}.
\end{equation}
For $\delta = +1$ the  variable $\tau = {\tau}(t)$ is monotonically
decreasing from $+ \infty $ to  $0$,  when $t-t_0$ is
varying from $0$ to $+ \infty $.
For $\delta = -1$ it is monotonically
decreasing from $\pi T$ to  $0$,  when $t-t_0$ is
varying from $ - \infty $ to $+ \infty $.

It is not difficult to verify that the following relations take place
\begin{eqnarray} 
\sinh(\tau \sqrt{\delta}/T)/ \sqrt{\delta} & =
& 1/{f_{\delta}}(q \sqrt{2{\cal E}}(t - t_{0})), \\
\tanh(\tau \sqrt{\delta}/2T)/ \sqrt{ \delta} & =
& \exp[-q \sqrt{2{\cal E}}(t - t_{0})],  \\
d\tau & = & - qT \sqrt{2{\cal E}} dt/ {f_{\delta}}(q \sqrt{2{\cal
E}}(t - t_{0})).
\end{eqnarray}

Now, we introduce the following dimensionless "Kasner-like"
parameters
\begin{eqnarray} 
&\beta^{i} = - e^{i}_{\hat{a}} p^{\hat{a}}/ (q \sqrt{ 2{\cal E}}), \\
&\beta_{\varphi} = - p^{n}/ (q \sqrt{ 2{\cal E}}),
\end{eqnarray}
Here and below $\hat{a}, \hat{b} =  1, \ldots, n -1$.
{}From relations (3.7), (3.14), (3.20) and (3.36) we have
\begin{eqnarray} 
x^i = & - (u^i/2q) z^0 +  e^{i}_{\hat{a}}
[p^{\hat{a}} (t - t_0) + \bar{q}^{\hat{a}}] =   \nonumber \\
 &- (u^i/4q^2) (2q z^0)  - q \sqrt{2{\cal E}}(t- t_{0}) \beta^{i} +
\gamma^{i},
\ea
where
\be  
\gamma^{i} = e^{i}_{\hat{a}} \bar{q}^{\hat{a}},
\qquad  \bar{q}^{\hat{a}} = q^{\hat{a}} + p^{\hat{a}}t_0.
\ee
Using (3.13), (3.27), (3.33), (3.34) and (3.38) we get the following
expression for the scale factors
\be 
a_{i} = \exp(x^i)= A_{i}[\sinh(r \tau /T)/r]^{\sigma^i}[\tanh (r \tau/
2T)/r]^{\beta^{i}},
\ee
where $r = \sqrt{\delta}$, and
\be 
\sigma^i  = 2 u^i/<u,u>_{*}, \qquad  A_{i} = ({\cal E}/|A|)^{\sigma^i/2}
\exp(\gamma^{i}),
\ee
$i=1, \ldots ,n$.  In analogous manner we obtain
the expression for the scalar field (see (3.34), (3.37))
\be 
\exp(\kappa \varphi) =  \exp(z^n) =
 A_{\varphi} [\tanh (r \tau/2T)/r]^{\beta_{\varphi}},
\ee
$A_{\varphi} > 0$ is constant.

We define a bilinear symmetrical form $<.,.>: R^{n} \times R^{n} \rightarrow
R$ by the relation
\be 
< \alpha ,\beta > = G_{ij} \alpha^i \beta^j ,
\ee
$\alpha = (\alpha^i)$, $\beta = (\beta^i)  \in R^{n}$.
Using the definitions (3.13), (3.22), (3.36), (3.37) and relations
(3.11), we obtain the relations between Kasner-like  parameters
\be 
< \beta, \beta > + (\beta_{\varphi})^2 = G_{ij} \beta^i \beta^j +
(\beta_{\varphi})^2 =
1/q^2 = - 4/<u,u>_{*},
\ee
and  (see (3.12))
\be  
u_{i} \beta^{i} = e^{0}_{i} e_{\hat{a}}^{i} P^{\hat{a}} =
\delta^{0}_{\hat{a}} P^{\hat{a}} = 0,
\ee
where $P^{\hat{a}} = - p^{\hat{a}} \sqrt{2/{\cal E}}$.

Analogously to
(3.45) we get $u_i \gamma^i = 0$ and hence (see (3.41))
\be 
\prod_{i=1}^{n} A_i^{u_{i}} =  {\cal E}/|A|.
\ee
Thus, the additional integral of motion ${\cal E}$ is a certain combination
of parameters $A_i$ and $|A|$ depending on the equation of state (2.8).

We introduce also a "quasi-volume" scale factor
\begin{equation} 
v = \prod_{i=1}^{n} a_{i}^{u_{i}/2} = \exp(\frac{1}{2} u_i x^i).
\end{equation}
{}From  (3.12), (3.27), (3.33), (3.46) (see also (3.40), (3.45) we have
\begin{equation} 
v =  v_0 \sinh(r \tau/T)/r = ({\cal E}/|A|)^{1/2}
({f_{\delta}}(q \sqrt{2{\cal E}}(t - t_{0})))^{-1},
\ee
Here
\be 
v_0 = \prod_{i=1}^{n} A_{i}^{u_{i}/2}.
\ee
The quasi-volume scale factor oscillates for $A < 0$
and exponentially increases  as $\tau \rightarrow +\infty$ for
$A > 0$.

{}From (3.3), (3.32), (3.35), (3.47), (3.48) we get
\ba 
\exp[2{\gamma_{0}}(t)] dt \otimes dt &= & (\prod_{i=1}^{n} a_{i}^{2N_{i}})
{f_{\delta}^2}(q \sqrt{2{\cal E}}(t - t_{0}))(2q^2 {\cal E} T^2)^{-1}
d\tau \otimes d\tau,  \nonumber \\
& = & (\prod_{i=1}^{n} a_{i}^{2N_{i}-u_{i}})  d \tau \otimes d\tau .
\ea

Thus we get the following solution  of the field equations (2.3) and (2.4)
\begin{eqnarray}  
&g=-(\prod_{i=1}^{n} ({a_{i}}(\tau))^{2N_{i}-u_{i}})
d\tau \otimes d\tau + \sum_{i=1}^{n} {a_{i}^{2}}(\tau)g^{(i)}, \\
&{a_{i}}(\tau) = A_{i}[\sinh(r \tau /T)/r]^{2 u^i/<u,u>_{*}}[\tanh (r \tau/
2T)/r]^{\beta^{i}},  \\
&\exp({\kappa \varphi}(\tau))=
A_{\varphi}  [\tanh (r\tau/2T)/r]^{\beta_{\varphi}}, \\
&\kappa^2 {\rho}(\tau) = A \prod_{i=1}^{n}
({a_{i}}(\tau))^{u_{i}-2N_{i}},
\end{eqnarray}
$i=1, \ldots ,n$; where $r
=\sqrt{A/|A|}$, $T$ is defined in (3.32),
$A_{i}, A_{\varphi}>0$  are constants
and the parameters $\beta^{i}, \beta_{\varphi}$ satisfy the relations
\be 
\sum_{i=1}^{n} u_{i} \beta^{i}=0,  \qquad
\sum_{i,j =1}^{n} G_{ij} \beta^{i} \beta^{j} +
(\beta_{\varphi})^2 = - 4/<u,u>_{*} = 1/q^2.
\ee
Here $\tau > 0$ for
$A > 0$  and $0 < \tau < \pi T$ for $A < 0$.

We note that the solution (3.51)-(3.55)  without scalar field
($\beta_{{\varphi}} = 0$) was obtained previously in
\cite{IM5}. For $u_i = 2 N_i$
($\Lambda$-term case) the solution was considered in \cite{BIMZ}
(for $\beta_{\varphi} = 0$ see also \cite{IM4}), where
Euclidean wormholes were constructed.

For small values of $\tau$ we have the following asymptotic
relations
\begin{equation}
{a_{i}}(\tau)   \sim   C_{i} \tau^{\bar{\beta}^i}, \qquad
\exp(\kappa {\varphi}(\tau)) \sim  C_{\varphi} \tau^{\beta_{\varphi}}
\end{equation}
as $\tau \rightarrow 0$, $i=1, \ldots ,n$, where
$C_{i}, C_{\varphi}$ are constants and
$\bar{\beta}^{i} = \beta^{i}   + \sigma^i$
are the new Kasner-like parameters, satisfying  the relations
\begin{equation}
u_{i} \bar{\beta}^{i} = 2, \qquad
G_{ij} \bar{\beta}^{i} \bar{\beta}^{j} + \beta_{\varphi}^2 = 0.
\end{equation}

\subsection{Exceptional solutions }

Now we consider the exceptional real solutions corresponding to
${\cal E} = 0$ and $A > 0$ (see (3.25)). From ${\cal E} = 0$ and
(3.22) we have $p^{B} = 0$  and hence
\be 
z^{B} = q^{B}
\ee
are constant, $B = 1, \ldots ,n$. So,
$\kappa \varphi = z^n = const$ in this case. From (3.7) and (3.12)
we have
\be 
x^i = - (u^i/4q^2) (2q z^0)  + \gamma^{i}, \qquad
\gamma^{i} = e^{i}_{\hat{a}} q^{\hat{a}},
\ee
($\hat{a} = 1, \ldots ,n -1$).  Using (3.13), (3.25), (3.32) and  (3.59)
for $t > t_0$, we get
\be  
a_{i} = \exp(x^i)=  [(t- t_0)/T]^{- \sigma^i} \exp(\gamma^{i}),
\ee
$i = 1, \ldots ,n$.

Introducing  the new time variable
\be   
T/(t- t_0) =   \exp[\pm (\tau - \tau_0)/T], \qquad t > t_0,
\ee
we  obtain
\begin{equation} 
{a_{i}}(\tau) =
\bar{A}_{i} \exp( \pm \sigma^i \tau /T),
\end{equation}
where
\be   
\bar{A}_{i}  = \exp( \mp \sigma^i \tau_0 /T) \exp(\gamma^{i}),
\ee
$i=1, \ldots ,n$.

Analogously to
(3.45) we get $u_i \gamma^i = 0$ and hence (see (3.63))
\be 
\prod_{i=1}^{n} \bar{A}_i^{u_{i}} =  \exp( \mp 2 \tau_0 /T).
\ee

For the quasi-volume  from (3.62) and (3.64) we get
\begin{equation} 
v = \prod_{i=1}^{n} a_{i}^{u_{i}/2} = \exp[ \pm (\tau - \tau_0)/T].
\ee

Thus, for  $A > 0$  we have a family of
exceptional solutions
with the constant real scalar field
\ba 
&g=-(\prod_{i=1}^{n} ({a_{i}}(\tau))^{2N_{i}-u_{i}}
                   )
d\tau \otimes d\tau + \sum_{i=1}^{n} {a_{i}^{2}}(\tau)g^{(i)}, \\
&{a_{i}}(\tau) = \bar{A}_{i} \exp[\pm 2 u^i \tau/(T <u,u>_{*})],  \\
&{\varphi}(\tau) = const,
\end{eqnarray}
and ${\rho}(\tau)$ is defined by (3.54).
Here $\bar{A}_{i} >0$ ($i=1, \ldots ,n$) are constants,
and $T$ is defined in (3.32).

We note that for $A > 0$ the
solution (3.67) with the sign $''+''$ is an attractor for the solutions
(3.52), i.e.
\begin{equation} 
{a_{i}}(\tau)   \sim   \bar{A}_{i} \exp( \sigma^i \tau/T),  \qquad
{\varphi}(\tau) \sim  const,
\end{equation}
$i=1, \ldots, n$, for $\tau \rightarrow +\infty$.

{\bf Synchronous-time parametrization.}
The relations (3.67) imply
\be 
\prod_{i=1}^{n} a_{i}^{2N_{i}-u_{i}} =
\bar{P}^2 \exp[ \pm 2 (\bar{\sigma} - 1) \tau /T],
\ee
where
\ba  
&\bar{P}  =  \prod_{i=1}^{n} \bar{A}_i^{N_i - \frac{1}{2}u_{i}}, \\
&\bar{\sigma} = 2N_i u^i/<u,u>_{*} = <u^{(\Lambda)}, u>_{*} /<u,u>_{*}.
\ea
Here and below
\be 
u^{(\Lambda)}_i = 2N_i.
\ee

Now, we introduce the synchronous time  variable $t_s$ satisfying the
relation
\be 
\bar{P}^2 \exp[ \pm 2 (\bar{\sigma} - 1) \tau /T] d\tau \otimes d\tau =
dt_s \otimes dt_s.
\ee

First, we consider the case
\be   
\bar{\sigma} \neq 1 \ \Longleftrightarrow \ <u^{(\Lambda)} - u, u>_{*}
\neq 0.
\ee
Introducing $t_s$ by the formula
\be   
t_s  =
\frac{\bar{P} T}{|\bar{\sigma} - 1|} \exp[ \pm (\bar{\sigma} - 1) \tau /T]
> 0 \ee we get for the scale factors
\be   
a_i = {a_i}(t_s)  = A_i t_s^{\nu^i},
\ee
where
\be 
\nu^i  = \sigma^i/(\bar{\sigma} - 1)  = 2
u^i/<u^{(\Lambda)} - u , u>_{*} \ee and
\be  
A_{i} = \bar{A}_{i}
[|\bar{\sigma} - 1|/(\bar{P} T)]^{\nu^i}.
\ee
The parameters $\nu^i$ (3.78) satisfy  the  following relation
\be  
\nu^i (2N_{i}-u_{i}) = 2.
\ee
The relations (3.32), (3.71), (3.79) and (3.80) imply
\be    
\prod_{i=1}^{n} A_{i}^{u_{i} - 2N_{i}} = T^2/(\bar{\sigma} - 1)^2
= - 2 < u , u>_{*}/ (A <u^{(\Lambda)} - u , u>_{*}^2).
\ee

{}From (3.54), (3.77), (3.80) and (3.81) we get the
following formula for the density
\be   
 \kappa^2 \rho = \kappa^2 {\rho}(t_s) = \frac{- 2 < u ,
 u>_{*}}{<u^{(\Lambda)} - u , u>_{*}^2 t_s^2}.
\ee

The  metric reads  as
 \be 
 g=- dt_s \otimes dt_s + \sum_{i=1}^{n}a_i^2(t_s) g^{(i)},
 \ee
where the scale factors are defined in (3.77), i=1, \ldots ,n.
Thus, formulas (3.54), (3.77), (3.82), (3.83) and $\varphi =
const$ describe exceptional solutions for the case (3.70). We call
these solutions as the power-law inflationary solutions.

Now we consider the case
\be  
\bar{\sigma} = 1 \ \Longleftrightarrow \ <u^{(\Lambda)} - u, u>_{*} = 0.
\ee
{}From (3.54), (3.70) we get
\be   
\kappa^2 \rho = A \bar{P}^{-2} = const.
\ee
Introducing synchronous time $t_s = \bar{P} \tau$, from (3.67)
we get
\begin{equation} 
{a_{i}}(t_s) =
\bar{A}_{i} \exp[ \mp \frac{u^i}{\sqrt{-<u,u>_{*}}} \frac{t_s}{T_0}],
\ee
where
\be 
T_0 = (2 \kappa^2 \rho)^{-1/2}.
\ee
The relations  (3.83), (3.85)-(3.87) and $\varphi = const$ describe
the exponential-type inflation for the case (3.84).

Let us consider the synchronous time parametrization for the solutions
(3.51)-(3.54). We have the following relation between
the synchronous time  $t_s$  and the $\tau$-variable
\be 
t_s = \varepsilon {F}(\tau), \qquad \frac{dF}{d \tau} = {f}(\tau),
\ee
where $\varepsilon = \pm 1$ and
\ba 
{f}(\tau) = \prod_{i=1}^{n} ({a_i}(\tau))^{N_{i} - \frac{1}{2} u_{i}}
= &&P [\sinh(r \tau /T)/r]^{\bar{\sigma} - 1} [\tanh (r \tau/
2T)/r]^{\beta^{i} N_i}, \\
\kappa^2 {\rho}(\tau) = &&A P^{-2} [\sinh(r \tau /T)/r]
^{2 - 2 \bar{\sigma}} [\tanh (r \tau/ 2T)/r]^{-2 \beta^{i} N_i},
\ea
with $P =  \prod_{i=1}^{n} A_i^{N_i - \frac{1}{2}u_{i}}$
and $\bar{\sigma}$ is  defined in (3.72).
{}From (3.90) we have
\be 
{f}(\tau) \sim B \tau^{p-1}, \qquad \tau \rightarrow +0,
\ee
where  $B > 0$ is constant and
\be 
p = {p}(\beta) = \bar{\sigma} + \beta^i N_i = (\sigma^i + \beta^i) N_i.
\ee
Putting  $\varepsilon = sign(p)$ from (3.88) and (3.91) we get
\be 
t_s  \sim  B_1 \tau^{p}  , \qquad \tau \rightarrow +0,
\ee
with $B_1 = B/|p|$ (here  the integration constant
in (3.88) is properly fixed).

{\bf Proposition 1.} Let $1/q^2 - (\beta_{\varphi})^2 \geq 0$.
Then, for all $\beta = (\beta^i)$ satisfying the relations
(3.55),  we have  ${p}(\beta) \neq 0$ and
\ba 
&&i) \  u^i N_i < 0  \ \Rightarrow  \ {p}(\beta) > 0, \\
&&ii) \ u^i N_i > 0  \ \Rightarrow  \ {p}(\beta) < 0.
\ea

The Proposition 1 is the special case of the more general
Proposition 2 proved in the Appendix.

{\bf Proposition 2.} Let two vectors $u = (u_i), v = (v_i)
\in R^n$  satisfy the inequalities:  \\
$<u,u>_{*} \equiv - 4 q^2 < 0$
and $<v,v>_{*} < 0$. Then,
$u^i v_i = <u,v>_{*} \neq 0$ and
for all $\beta = (\beta^i)$ satisfying the relations
\be 
u_i \beta^{i} = 0,  \qquad
G_{ij} \beta^{i} \beta^{j}  \leq 1/q^2,
\ee
the following relation is valid
\be 
sign(u^i v_i) = - sign ((\sigma^i + \beta^i) v_i),
\ee
where $\sigma^i  = 2 u^i/<u,u>_{*}$.

For the vector
\be 
v_i = N_i = \frac{1}{2} u^{(\Lambda)}_i
\ee
we have
\be 
v^i = N^i =  G^{ij} N_{j} = \frac{1}{2 - D}
\ee
and hence
\be 
<v,v>_{*} = N_i N^i =  - \frac{D - 1}{D - 2} < 0.
\ee
Thus the relations (3.92), (3.100) and the Proposition 2 imply
the Proposition 1.

{}From  (3.99) we get
\be 
u_i N^i =   \frac{1}{2 - D}  \sum_{i=1}^{n} u_i.
\ee
Using relation (3.93), (3.101) and Proposition 1
we obtain
\be 
t_s  \rightarrow + 0, \qquad as \ \tau \rightarrow + 0,
\ee
for
\be 
A) \ \sum_{i=1}^{n} u_i > 0, \qquad {p}(\beta) > 0
\ee
and
\be 
t_s  \rightarrow + \infty, \qquad as \ \tau \rightarrow + 0,
\ee
for
\be 
B) \ \sum_{i=1}^{n} u_i < 0, \qquad {p}(\beta) < 0.
\ee

In the limit $\tau \rightarrow + 0$ we have
$\tau \sim  (t_s/B_1)^{1/p}$ (see (3.93)) and hence (see (3.56))
\be 
{a_i}(t_s)  \sim  \bar{B_i} t_s^{\alpha^i},  \
\exp(\kappa {\varphi}(t_s)) \sim
\bar{B}_{\varphi} t_s^{\alpha_{\varphi}}
\ee
as $t_s  \rightarrow + 0$ in the case A) (3.103) and
as $t_s  \rightarrow + \infty$ in the case B) (3.105).
Here $\bar{B_i}, \bar{B}_{\varphi}$ are constants and
\be 
\alpha^i =
(\sigma^i + \beta^i)/{p}(\beta), \
\alpha_{\varphi} = \beta_{\varphi}/{p}(\beta),
\ee
$i=1 \ldots n$.
The parameters $\alpha^i, \alpha_{\varphi}$  satisfy the Kasner-like
relations
\begin{equation}
\sum_{i=1}^{n} N_{i} \alpha^{i} = 1,
\ee
\be 
\sum_{i=1}^{n} N_{i} (\alpha^{i})^2  + \alpha_{\varphi}^2 = 1.
\ee

The first relation (3.108) is quite obvious, the second (3.109) is
following from the first one,  (2.12) and the relation
\be 
G_{ij} \alpha^{i} \alpha^{j} + \alpha_{\varphi}^2 = 0,
\ee
that can be readily verified using the relations (3.13), (3.41),
(3.55) and (3.107).

The Kasner-like asymptotical behaviour (3.106), (3.108), (3.109)
for the case A)
agrees with one of the results of  \cite{IM6}:  in the case A) the
perfect fluid components with $<u,u>_{*} < 0$ may be neglected near the
singularity $t_s \rightarrow + 0$ and we are lead to the Kasner-like
formulas \cite{BZ3} (see also \cite{I1}).

\begin{center}     Fig. 1               \end{center}

We note that for the case $n = 2$
the following relation takes place:
\be 
[b_2 \frac{u_1}{N_1} - (1 + s) \frac{u_2}{N_2}]
[b_1 \frac{u_2}{N_2} - (1 + s) \frac{u_1}{N_1}] =
= (- s^2) (1 + s) <u,u>_{*},
\ee
where $b_i = 1 - \frac{1}{N_i}$, $i =1,2$ and $s = \sqrt{1 - b_1 b_2}$.
This implies the relations for the light-cone lines ($<u,u>_{*} = 0$):
\ba 
l_1: \qquad b_2 (1 - \xi_1) = (1 + s) (1 - \xi_2) ,    \\
l_2: \qquad b_1 (1 - \xi_2) = (1 + s) (1 - \xi_1),
\ea
(see Fig.1).

\subsection{Isotropization-like behaviour}

Here we rewrite the attractor behaviour (3.69)
for the non-exceptional solutions (3.51)-(3.55) with $A > 0$
(as $\tau \rightarrow +\infty$) in terms of the synchronous
time variable $t_s$.
For the function (3.89) we have the following asymptotical behaviour
\be 
{f}(\tau) \sim
P [\frac{1}{2} \exp( \tau /T)]^{\bar{\sigma} - 1} =
\bar{B} \exp[(\bar{\sigma} - 1) \tau /T],
\ee
when $\tau \rightarrow +\infty$ ($\bar{B} = const$).

First we consider the case $\bar{\sigma} = 1$ (see (3.84)).
In this case ${f}(\tau) \sim \bar{B}$
as $\tau \rightarrow +\infty$
and hence (see (3.88))
\be 
t_s = {F}(\tau) \sim \bar{B} \tau + C,
\ee
as $\tau \rightarrow + \infty$.
(Due to  $ u^i N_i < 0 $ and Proposition 1  $\varepsilon = sign(p) = +1$).
The synchronous time $t_s$ is
monotonically increasing from $0$ to $+ \infty$ as  $\tau$
is varying  from $0$ to $+ \infty$ (see (3.93)).

{}From (3.54), (3.69) and  (3.115)  for the case $\bar{\sigma} = 1$
we get
\ba   
&&{a_{i}}(t_s) \sim
\bar{A}_{i} \exp[ - \frac{u^i}{\sqrt{-<u,u>_{*}}} \frac{t_s}{T_0}], \\
&&{\varphi}(t_s) \sim  const, \\
&&{\rho}(t_s) \sim \rho_0.
\ea
when $t_s \rightarrow +\infty$,
where $T_0 = (2 \kappa^2 \rho_0)^{-1/2}$.

Now, we consider the case   $\bar{\sigma} \neq 1$ (see (3.75)).
In this case
\be   
{F}(\tau) \sim
\frac{\bar{B} T}{(\bar{\sigma} - 1)} \exp[(\bar{\sigma} - 1) \tau /T]
+ C,
\ee
where $C$ is constant.

First, we consider the subcase:  $\bar{\sigma} > 0$  or, equivalently,
$u_i N^i < 0$  (or $\sum_{i=1}^{n} u_i > 0$, see (3.101)).
We have $t_s =  {F}(\tau)$, since $p > 0$ due to (3.103)
and $\varepsilon = sign (p) = + 1$).
In this case $t_s$ is monotonically increasing from $0$
to $T_{*} > 0$ for  $ 0 < \bar{\sigma} < 1$
and
to $+ \infty$ for  $\bar{\sigma} > 1$
as  $\tau$ is varying  from $0$ to $+ \infty$
(see (3.93)). Using (3.54), (3.69) we get
\ba   
 &&{a_i}(t_s)  \sim  A_i (T_{*} -t_s)^{\nu^i}, \\
 &&{\varphi}(t_s) \sim  const, \\
 &&\kappa^2 {\rho}(t_s) \sim \frac{- 2 < u ,
 u>_{*}}{<u^{(\Lambda)} - u , u>_{*}^2 (T_{*} -t_s)^2}.
\ea
as $t_s \rightarrow T_{*} - 0$,
for  $\bar{\sigma} < 1$.
For  $\bar{\sigma} > 1$ we have the asymptotic behaviour
in the limit $t_s \rightarrow + \infty$ described by the relations
\ba   
&&{a_i}(t_s)  \sim  A_i t_s^{\nu^i}, \\
&&{\varphi}(t_s) \sim  const, \\
&&\kappa^2 {\rho}(t_s) \sim \frac{- 2 < u ,
u>_{*}}{<u^{(\Lambda)} - u , u>_{*}^2 t_s^2}.
\ea
as $t_s \rightarrow + \infty$, where $\nu^i$ is defined in (3.78).

Now, we consider the subcase:  $\bar{\sigma} < 0$  or, equivalently,
$u_i N^i > 0$  (or $\sum_{i=1}^{n} u_i < 0$, see (3.101))
We remind that $p < 0$ due to (3.105) and $\varepsilon = sign (p) = -1$
. Then, we have $t_s = - {F}(\tau)$  (we put
$C = 0$ in (3.119)) and $t_s$ is monotonically decreasing from
$+ \infty$  to $0$ as  $\tau$ is varying  from $0$ to $+ \infty$
(see (3.93)). In the considered subcase we have the asymptotic
behaviour in the limit $t_s \rightarrow + 0$ described by
the relations (3.123)-(3.125).

\subsection{Solutions with pure imaginary scalar field}

Now, we consider the solutions of the field equations with the complex
scalar field and the real metric. In this case
${{{\cal E}}}, p^{1}, \ldots, p^{n-1}$ are real and hence (see
(3.21), (3.22))
$p^{n}$  is either real or pure imaginary. The case of real $p^{n}$ was
considered above.

For the pure imaginary $p^{n}$ we have three subcases:
a) ${{{\cal E}}} >0$, b) ${{{\cal E}}} = 0$, c) ${{{\cal E}}} < 0$.
In the first case  ${{{\cal E}}} > 0$
after the reparametrization (3.29)-(3.32) we get the solutions
(3.51)-(3.55) with an imaginary value of
$\beta_{\varphi}$. The cases  b) and c): ${{{\cal E}}} \leq 0$ take
place only for $A > 0$ (see (3.24), (3.25)).

Let us consider the case  ${{\cal E}} < 0$.
Here, we have (see (3.36), (3.37)) the  imaginary
${\beta}^{k}$:
\begin{equation}  
{\beta}^{k} = i \hat{\beta}^{k}, \qquad k =1, \ldots, n,
\end{equation}
and real  $\beta_{\varphi}$.
The solution may be obtained from
(3.51)-(3.55) substituting (3.126) and $\tau/T \mapsto \tau/T
+ i \frac{\pi}{2}$:
\begin{eqnarray}  
g &=&- (\prod_{i=1}^{n} ({a_{i}}(\tau))^{2N_{i}-u_{i}})
d\tau \otimes d \tau + \sum_{i=1}^{n} {a_{i}^{2}}(\tau) g_{(i)}, \\
{a_{i}}(\tau) &=& \hat{A}_{i}[\cosh(\tau/T)]^{\sigma^i}
[{f}(\tau/2T)]^{\hat{\beta}^{i}}, \\
{\varphi}(\tau) &=& c + 2i
\beta_{\varphi} \arctan \exp(- \tau/T),  \end{eqnarray}
where $c, \hat{A}_{i} \neq 0$ are constants, $i =1, \ldots ,n$,
$T$ is defined in (3.32), $\sigma^i$ are defined in (3.41),
 $A > 0$ and the real
parameters $\hat{\beta}^{i}, \beta_{\varphi}$ satisfy the relations
\be 
\sum_{i=1}^{n} u_{i} \hat{\beta}^{i}= 0,  \qquad
- \sum_{i,j =1}^{n} G_{ij} \hat{\beta^{i}} \hat{\beta^{j}} +
(\beta_{\varphi})^2 = - 4/<u,u>_{*} = 1/q^2.
\ee
Here, like in \cite{BIMZ},
\begin{equation} 
{f}(x) \equiv
[\tanh(x + i \frac{\pi}{4})]^{i} = \exp(-2 \arctan e^{-2x})
\end{equation}
is the smooth monotonically increasing function bounded by
its asymptotics: \hfil  \\$e^{-\pi} < {f}(x) < 1$;
${f}(x) \rightarrow 1$ as $x
\rightarrow + \infty$ and  ${f}(x) \rightarrow e^{-\pi}$ as $x
\rightarrow - \infty$.
The solution (3.127)-(3.130) (with $\rho$ from (3.54))
may be also obtained from formulas (3.20), (3.24).  The
relation between the harmonic time and $\tau$-variable
(3.33) for the  case ${{{\cal E}}} < 0$  is modified
\begin{equation} 
\cosh(\tau /T)= 1/ \sin(q \sqrt{2|{\cal E}|}(t - t_{0})).
\end{equation}
For the quasi-volume scale factor we have
\begin{equation} 
v = \prod_{i=1}^{n} a_{i}^{u_{i}/2} =
(\prod_{i=1}^{n} \hat{A}_{i}^{u_{i}/2}) \cosh(\tau /T).
\end{equation}
The scalar field  ${\varphi}(t)$ varies from $c +  i \pi
\beta_{\varphi}$ to $c$ as  $\tau$  varies from $- \infty$  to $+ \infty$.
The solution (3.127)-(3.130)
for $\tau \in (-\infty ,+ \infty)$ is non-singular .
Any scale factor ${a_{i}}(\tau)$ for some $\tau_{0i}$ has a minimum
and
\begin{equation}  
{a_{i}}(\tau) \sim A_{i}^{\pm}
\exp(\sigma^i |\tau| /T),
\end{equation}
for $\tau \rightarrow \pm \infty$.

The "Lorentzian" solutions considered above have
"Euclidean" analogues for $A < 0$ also
\begin{eqnarray}  
&&g=
(\prod_{i=1}^{n} ({a_{i}}(\tau))^{2N_{i}-u_{i}})
d\tau \otimes d\tau + \sum_{i=1}^{n} {a_{i}^{2}}(\tau) g_{(i)}, \\
&&{a_{i}}(\tau) = \hat{A}_{i}[\cosh(\tau /T)]^{\sigma^i}
[{f}(\tau/2T)]^{\hat{\beta}^{i}},   \\
&&{\varphi}(\tau) = c + 2i \sigma_{n+1} \arctan \exp(- \tau/T),
\end{eqnarray}
with the parameters $\hat{\beta}^{i}, \beta_{\varphi}$
satisfying  the relations
(3.130).
When all spaces $(M_{i}, g^{(i)})$ are Riemannian,
this solution may be interpreted
as the classical Euclidean wormhole (instanton) solution.

An interesting  special case of the solution (3.135)-(3.137) occurs for
$\hat{\beta}^{i} = 0$, $i =1, \ldots ,n$, (this corresponds
to $p^{\hat{a}} = 0$)
\begin{eqnarray}  
{a_{i}}(\tau) &=& \hat{A}_{i}[\cosh(\tau /T)]^{\sigma^i}, \\
{\varphi}(\tau) &=& c \pm 2i q^{-1} \arctan \exp(- \tau/T).
\end{eqnarray}
All scale factors (3.138) have a minimum at $\tau = 0$ and
are symmetric with respect to the time inversion:
$\tau \mapsto - \tau$.
It is necessary to stress that here, like in \cite{BIMZ}, wormhole
solutions take place only in the presence of the  imaginary scalar field.

\subsection{Some examples}
In this subsection we consider some application of the obtained
above formulas.

{\bf 3.5.1. Isotropic case.}

First, we consider the isotropic case:
\be 
u_i = h N_i \Longleftrightarrow \ u = \frac{h}{2} u^{(\Lambda)},
\qquad p_i = (1 - h) \rho,
\ee
where $h \neq 0$ is constant.  From (3.98)-(3.100) and (3.140) we
get
\be 
u^i = \frac{h}{2 - D}, \qquad
<u,u>_{*} =  - h^2 \frac{D - 1}{D - 2} < 0,
\ee
and hence
\be 
\sigma^i  = 2 u^i/<u,u>_{*} = \frac{2}{h(D-1)} = {\sigma}(h) = \sigma.
\ee

The solution (3.51)-(3.55) reads

\begin{eqnarray}  
&g=-(\prod_{i=1}^{n} ({a_i}(\tau))^{(2-h) N_{i}})
d\tau \otimes d\tau + \sum_{i=1}^{n} {a_i^2}(\tau) g^{(i)}, \\
&{a_{i}}(\tau) = A_{i}[\sinh(r \tau /T)/r]^{{\sigma}(h)}[\tanh (r \tau/
2T)/r]^{\beta^{i}},  \\
&\exp({\kappa \varphi}(\tau))=
A_{\varphi}  [\tanh (r\tau/2T)/r]^{\beta_{\varphi}}, \\
&\kappa^2 {\rho}(\tau) =
A \prod_{i=1}^{n}({a_{i}}(\tau))^{(h-2)N_{i}} =
A (\prod_{i=1}^{n} A_{i}^{(h-2)N_{i}}) [\sinh(r \tau /T)/r]^{2(h-2)/h},
\end{eqnarray}
$i=1, \ldots ,n$; where $r
=\sqrt{A/|A|}$,
\be 
T = |h|^{-1} \left[ \frac{|A|(D - 1)}{2(D - 2)} \right]^{-1/2},
\ee
$A_{i}, A_{\varphi}>0$  are constants
and the parameters $\beta^{i}, \beta_{\varphi}$ satisfy the relations
\be 
\sum_{i=1}^{n} N_{i} \beta^{i}= 0,  \qquad
\sum_{i=1}^{n} N_{i} (\beta^{i})^{2} +
(\beta_{\varphi})^2 = \frac{4(D - 2)}{h^2 (D - 1)}.
\ee
The special case of this solution with $h =2$ ($\Lambda$-term case)
was considered in \cite{BIMZ}.

Now, we consider the exceptional solutions for $A > 0$. From
(3.72) and (3.141) we have
\be 
\bar{\sigma} = \sigma^i N_i = 2/h,
\qquad <u^{(\Lambda)} - u, u>_{*} =
h(h-2) \frac{D - 1}{D - 2}.
\ee
{}From (3.149) we get: $<u^{(\Lambda)} - u, u>_{*} =0 \
\Longleftrightarrow \ h = 2$ ($h \neq 0$).
The matter corresponds to cosmological constant
$\Lambda = \kappa^2 \rho > 0$. The relations (3.86), (3.141)
imply the solution \cite{IM4,BIMZ} with metric (3.83) and
\be 
{a_{i}}(t_s) =
\bar{A}_{i} \exp[\pm \frac{t_s \sqrt{2 \Lambda}}{\sqrt{(D-1)(D-2)}}]
\ee
($\varphi = const$).

For $h \neq 2$ ( $\Longleftrightarrow \ <u^{(\Lambda)} - u, u>_{*} \neq
0$)
\be 
\nu^i  = 2 u^i/<u^{(\Lambda)} - u , u>_{*} =
\frac{2}{(2-h)(D-1)} = {\nu}(h) = \nu.
\ee
{}From (3.77), (3.82), (3.141), (3.149)
and (3.151)  we obtain the relations for scale factors and density:
\ba
&&{a_i}(t_s)  = A_i t_s^{{\nu}(h)},  \\
&&\kappa^2 {\rho}(t_s)
= \frac{2 (D-2)}{(h-2)^2 (D-1) t_s^2}.
\ea
For $h < 2$ (or $p > - \rho$)
we have an isotropic expansion of all scale factors and for $h > 2$ (or $p
< - \rho$) we have an isotropic contraction (see Fig.1).

{\bf Kasner-like behaviour.} In the considered case
$\sum_{i=1}^{n} u_i = h (D-1)$ and hence
(see (3.102)-(3.108)) the Kasner-like behaviour (3.106),(3.108), (3.109)
takes place as:
A) $t_s  \rightarrow + 0$, for
for $h > 0$ (or $p < \rho$), and
B) $t_s  \rightarrow + \infty$, for
for $h < 0$ (or $p > \rho$).

{\bf Isotropization-like behaviour.} Using the results of subsection
3.3 and (3.149)
we are lead to the following attractor behaviour:
\ba 
&&{a_i}(t_s)  \sim A_i t_s^{{\nu}(h)},  \\
&&\kappa^2 {\rho}(t_s) \sim \frac{2 (D-2)}{(h-2)^2 (D-1) t_s^2}.
\ea
in the limits
$t_s  \rightarrow + \infty$,
for $ 0 < h < 2$ (or $ - \rho < p <  \rho$)
and  $t_s  \rightarrow + 0$,
for $ h < 0$ (or $ p > \rho$).

Remark 2. For the dust matter case $h = 1$  ($p = 0$),
$\rho > 0$, the solution  (3.143)-(3.148)
has the sinchronous-time representation:
\begin{eqnarray}  
&g= - dt_s \otimes dt_s + \sum_{i=1}^{n} {a_{i}^{2}}(t_s)g^{(i)},
\nonumber \\
&{a_{i}}(t_s) = \bar{A}_{i}[t_s(t_s + T_1)]^{1/(D-1)}
[t_s/(t_s + T_1)]^{\beta^{i}/2},  \\
&\exp(2 {\kappa \varphi}(t_s))= A_{\varphi}
[t_s/(t_s + T_1)]^{\beta_{\varphi}}, \\
&\kappa^2 {\rho}(t_s) =
2 (D-2)/[(D-1) t_s(t_s + T_1)].
\end{eqnarray}
$i=1, \ldots ,n$; where $0 < t_s < + \infty$,
$T_1 > 0$, $\bar{A}_{i}, A_{\varphi}>0$
are constants and the
parameters $\beta^{i}, \beta_{\varphi}$ satisfy the relations
\be 
\sum_{i=1}^{n} N_{i} \beta^{i}= 0,  \qquad
\sum_{i=1}^{n} N_{i} (\beta^{i})^{2} +
(\beta_{\varphi})^2 = \frac{4(D - 2)}{(D - 1)}.
\ee
The special case of this solution with $\beta_{\varphi} = 0$ was
considered previously in \cite{IM2} (for $n = 2$ and $N_1 = \ldots N_n$
see \cite{Lor} and \cite{BO} correspondingly.)

{\bf 3.5.2 Curvature-like component.}

Now we consider the perfect fluid matter
with
\be 
u_i = 2 h ( - \delta_i^1 +  N_i) = h u^{(1)}_i,
\ee
where
$h \neq 0$ is constant and $N_1 > 1$. For $h =1$ this component
corresponds to the curvature term for the first space \cite{IMZ} (see
below).  The calculation gives
\be 
u^i = - \frac{2 h}{N_1} \delta^i_1,  \qquad
<u,u>_{*} =  - 4 h^2 b_1 < 0,
\ee
where $b_1 = 1- \frac{1}{N_1}$ and
\be 
<u, u^{(\Lambda)}>_{*} = 2 u^i N_{i} = -4h,
\ \
<u, u^{(\Lambda)} - u>_{*} = 4h (-1 + h b_1),
\ \
\sigma^i  = \frac{\delta^i_1}{h(N_1-1)}.
\ee
Using (3.160) and (3.162) we get from (3.51)-(3.55):
\begin{eqnarray}  
&g= - ({a_{1}}(\tau))^{2h}
(\prod_{i=1}^{n}({a_{i}}(\tau))^{2N_{i}(1-h)})
d\tau \otimes d\tau + \sum_{i=1}^{n} {a_{i}^{2}}(\tau)g^{(i)}, \\
&{a_{1}}(\tau) = A_{1}[\sinh(r \tau /T)/r]^{\frac{1}{h(N_1 -1)}}
[\tanh (r \tau/ 2T)/r]^{\beta^{1}},  \\
&{a_{i}}(\tau) = A_{i}[\tanh (r \tau/2T)/r]^{\beta^{i}}, \ \ i > 1, \\
&\exp({\kappa \varphi}(\tau))=
A_{\varphi}  [\tanh (r\tau/2T)/r]^{\beta_{\varphi}}, \\
&\kappa^2 {\rho}(\tau) = A ({a_{1}}(\tau))^{-2h}
(\prod_{i=1}^{n}({a_{i}}(\tau))^{2N_{i}(h - 1)})
\end{eqnarray}
$i=1, \ldots ,n$; where $r
= \sqrt{A/|A|}$, $T = |h|^{-1} (2|A| b_1)^{-1/2}$,
$A_{i}, A_{\varphi} > 0$  are constants
and the parameters $\beta^{i}, \beta_{\varphi}$ satisfy the relations
\be 
\beta^{1} = \frac{1}{1 - N_1} \sum_{i= 2}^{n} N_{i} \beta^{i},
\qquad
(\sum_{i= 2}^{n} N_{i} \beta^{i})^2  +
(N_1 -1) [\sum_{i= 2}^{n} N_{i} (\beta^{i})^2 + (\beta_{\varphi})^2] =
N_1 h^{-2}.
\ee

For $h \neq h_0 \equiv b_1^{-1} = N_1/(N_1 - 1) > 1$ we have from
(3.162) $<u, u^{(\Lambda)} - u>_{*} \neq 0$ and (see (3.78))
\be      
\nu^i  =  \delta^i_1 {\nu}(h), \qquad {\nu}(h) =
[N_1 + h (1 - N_1)]^{-1}.
\ee
The power-law inflationary solution for the considered
case reads:
\ba    
&&g=- dt_s \otimes dt_s + A_1^2 t_s^{2{\nu}(h)} g^{(1)} +
\sum_{i=2}^{n} A_i^2 g^{(i)},   \\
&& \varphi = const,      \\
&&\kappa^2 {\rho}(t_s) = \frac{b_1}{2 (-1 + h b_1)^2  t_s^2}.
\ea
The scale factors of internal spaces in this solution are
constant (we have the so-called "spontaneous compactification").
It is note difficult to show that  constancy of internal
scale factors leads to the equation of state (3.160).

Using the  relation $\bar{\sigma} = h_0/h$ and the analysis
carried out in subsection 3.3 we obtain
that the solution (3.169)-(3.172)
is an attractor for non-exceptional solutions with $\rho > 0$ as
 $t_s  \rightarrow T_{*} - 0$, for $h > h_0$;
 $t_s  \rightarrow + \infty$, for $ 0 < h < h_0$
and  $t_s  \rightarrow + 0$, for $ h < 0$. So, we obtained
the solutions with the  "dynamical compactification".

{\bf 1-curvature case.} Here we apply the obtained above relations
to the cosmological model described by the action
\begin{equation}   
S = \int d^{D}x \sqrt{|g|} \{
 \frac{1}{\kappa^2} {R}[g] -  \partial_{M} \varphi
 \partial_{N} \varphi g^{MN} \}
\ee
with a scalar field $\varphi = {\varphi}(t)$ and
metric  (2.1) defined on the manifold (2.2), where
$(M_{i}, g^{(i)})$, $i = 2, \ldots ,n$; are Ricci-flat spaces
and $(M_{1}, g^{(1)})$  is an Einstein space
of non-zero curvature, i.e. ${R_{mn}}[g^{(1)}] =
 \lambda^1 g^{(1)}_{mn}, \lambda^1 \neq 0$.
Here $n \geq2$ and $N_{i} = dim M_i$. This "1-curvature model"
is equivalent to a special case of the considered above model
(3.160) with $h = 1$  and $A = - \frac{1}{2} \lambda^1 N_1$.
(see \cite{I1}). The solution (3.163)-(3.168) reads for this case
\begin{eqnarray}  
 &g= ({a_{1}}(\tau))^{2} [ - d\tau \otimes d\tau  + g^{(1)}]
 + \sum_{i=2}^{n} {a_{i}^{2}}(\tau)g^{(i)}, \\
 &{a_{1}}(\tau) = A_{1}[\sinh(r \tau /T)/r]^{\frac{1}{(N_1 -1)}}
 [\tanh (r \tau/ 2T)/r]^{\beta^{1}},  \\
 &{a_{i}}(\tau) = A_{i}[\tanh (r \tau/2T)/r]^{\beta^{i}}, \  i > 1, \\
 &\exp({\kappa \varphi}(\tau))=
 A_{\varphi}  [\tanh (r\tau/2T)/r]^{\beta_{\varphi}}, \\
 &\kappa^2 {\rho}(\tau) = A ({a_{1}}(\tau))^{-2}
 \end{eqnarray}
$i=1, \ldots ,n$; where $r
=\sqrt{- \lambda^1/|\lambda^1|}$, $T = [|\lambda^1| (N_1 - 1)]^{-1/2}$,
$A_{i}, A_{\varphi}>0$  are constants
and the parameters $\beta^{i}, \beta_{\varphi}$ satisfy the relations
\be 
\beta^{1} = \frac{1}{1 - N_1} \sum_{i= 2}^{n} N_{i} \beta^{i},
\qquad
\frac{1}{1 - N_1} (\sum_{i= 2}^{n} N_{i} \beta^{i})^2  +
\sum_{i= 2}^{n} N_{i} (\beta^{i})^2  +  (\beta_{\varphi})^2 =
\frac{N_1}{(N_1 -1)}.
\ee

The power-law inflationary solution for the negative curvature
case $\lambda^1 < 0$ reads:
\ba    
&&g=- dt_s \otimes dt_s + A_1^2 t_s^{2} g^{(1)} +
\sum_{i=2}^{n} A_i^2 g^{(i)},   \\
&& \varphi = const,
\ea
where $A_1^2 = |\lambda^1|/(N_1 - 1)$ (see (3.81), (3.172)). We are lead
here to the Milne-type solution recently considered in
\cite{BZ2}.

There exist another parametrization of the solution (3.174)-
(3.179) in terms of $R$-variable related to $\tau$-variable
as
 \ba 
 F = {F}(R) && = 1 - (\frac{R_0}{R})^{N_1 -1} = \tanh^2(\tau/2T),
 \qquad \lambda^1 < 0,  \\
 && = (\frac{R_0}{R})^{N_1 - 1} - 1 = \tan^2(\tau/2T),
 \qquad \lambda^1 > 0.
 \ea
Here $R > R_0$ for $\lambda^1 < 0$
and  $R < R_0$ for $\lambda^1 > 0$;
 $R_0 = A_1 2^{1/(N_1-1)} \sqrt{(N_1-1)/|\lambda^1|}$.
In new variables the metric and the scalar field may be written
as
 \ba 
 &g= - F^{b-1}  dR \otimes dR  + F^{b} R^2
 A_1^2 g^{(1)}
 + \sum_{i=2}^{n} F^{\beta^i} A_i^2 g^{(i)}, \\
 &\exp(2 \kappa {\varphi}(R))=
 A_{\varphi}^2  F^{\beta_{\varphi}},
 \end{eqnarray}
$A_1^2 = |\lambda^1|/(N_1 - 1)$, $A_{i}, A_{\varphi}>0$  are constants
and
 \be 
   b = (1 - \sum_{i= 2}^{n} N_{i} \beta^{i})/(N_1 - 1) =
  (N_1 - 1)^{-1} + \beta^{1},
 \ee
and the parameters  $\beta^{i} (i >1), \beta_{\varphi}$ satisfy the
relations (3.179). The  special case of the solution (3.179),
(3.184)-(3.186)  with $\beta_{\varphi} =0$ (constant scalar field)
was obtained earlier in \cite{I1}.

Remark 3. As a special case  of the presented above solution we
get a scalar-vacuum analogue of
spherically-symmetric Tangherlini solution \cite{Tan}
with  $n$  Ricci-flat internal spaces:
 \begin{eqnarray}
 g= &&- f^{a} dt \otimes dt + f^{b-1}  dR \otimes dR
 + f^{b} R^{2} d \Omega^{2}_{d} +  \sum_{i=1}^{n} f^{a_{i}}
 B_i g^{(i)}, \\
 &&\exp(2 {\kappa \varphi}(R))= B_{\varphi}  f^{a_{\varphi}},
 \end{eqnarray}
where $d \Omega^{2}_{d}$
is the canonical metric on $d$-dimensional sphere
 $S^{d}$ ($d \geq 2$),  $f = {f}(R) = 1 - B R^{1-d}$,
 $B_{\varphi}, B_i > 0,
B$ are  constants
and the parameters $a, a_{1}, \ldots , a_{n}$ satisfy
the relation
 \ba
 b = (1 - a - \sum_{i=1}^{n} a_{i}N_{i})/(d-1),  \\
 (a + \sum_{i=1}^{n} a_{i}N_{i})^{2} +
 (d-1) (a^{2} + a_{\varphi}^{2} +
 \sum_{i=1}^{n} a_{i}^{2} N_{i})= d.
\ea
For $a_{\varphi} = 0$ see also \cite{FIM2,IM8}. In the parametrization
of the harmonic-type variable this solution was presented earlier
in \cite{BM,IM8}.


\section{Wheeler-DeWitt equation}
\setcounter{equation}{0}


The quantization of the zero-energy constraint (3.17) leads to the
Wheeler-DeWitt (WDW) equation in the harmonic time gauge (3.3)
\cite{IMZ,IM5,BIMZ}
\begin{equation} 
2 \hat H \Psi \equiv
\left[ \frac{\partial}{\partial z^{0}} \frac{\partial}{\partial z^{0}}
- \sum_{i=1}^{n} \frac{\partial}{\partial z^{i}}
\frac{\partial}{\partial z^{i}} + 2 A \exp(2qz^{0}) \right] \Psi = 0.
\end{equation}

We are seeking the solution of (4.1) in the form
\be 
{\Psi}(z)=\exp(i \vec{p} \vec{z}){\Phi}(z^{0}),
\end{equation}
where $\vec{p} = (p^{1}, \ldots , p^{n})$ is a constant vector
(generally from C$^{n}$), $\vec{z}=(z^{1}, \ldots, z^{n-1}, z^{n}=
\kappa \varphi )$,
$\vec{p}\vec{z} \equiv \sum_{i=1}^n p_{i}z^{i}$.
Substitution of (4.2) into (4.1) gives
\be 
[- \frac12 (\frac{\partial}{\partial z^{0}})^{2} + V_0(z^{0})]
\Phi = {\cal E} \Phi,
\end{equation}
where
${\cal E}= \frac{1}{2} \vec{p}\vec{p}$ and
$V_0(z^0) = - A \exp(2qz^0)$.
Solving (4.3), we get
\be 
{\Phi}(z^{0})={B_{i\sqrt{2{{\cal E}}}/q}}(\sqrt{-2 A}
q^{-1} \exp(qz^{0})),
\end{equation}
where $i \sqrt{2{{\cal E}}}/q= i|\vec{p}|/q,$ and $B=I, K$
are modified Bessel functions.
We note, that
\be 
v = \exp(qz^{0})= \prod_{i=1}^{n}a_{i}^{u_{i}/2}
\end{equation}
is the "quasi-volume" (3.47) (see (3.12)).

The general solution of Eq. (4.1) has the following form
\be 
{\Psi}(z) = \sum_{B=I,K} \int d^n \vec{p}
{~C_{B}}(\vec{p}) e^{i\vec{p}\vec{z}}{B_{i|\vec{p}|/q}}
(\sqrt{-2 A} q^{-1} \exp(qz^{0})),
\end{equation}
where functions $C_{B}$ ($B=I,K$) belong to an
appropriate class.
For the $\Lambda$-term case this solution was considered in
\cite{IM4,BIMZ} and for the two-component model  ($n=2$) and
$\Lambda > 0$ in \cite{GKag}.

In the ground state we put all momenta $p^a (a =
1, \dots, n)$ equal to zero and the ground state wave function reads
\be 
\Psi_0 = B_0 \left(\sqrt{-2 A} q^{-1} \exp(qz^0) \right).
\ee
The function  $\Psi_0$ is invariant with respect to
the rotation group O(n).

Remark 4.
Applying the arguments considered in \cite{Zh1,BIMZ} one may  show
that the  ground state wave function
\ba 
\Psi_0^{(HH)} =
&I_0 \left( \frac{\sqrt{2|A|}}{q} \exp(qz^0) \right) , \qquad A < 0, \\
&J_0 \left( \frac{\sqrt{2A}}{q} \exp(qz^0) \right),  \qquad  A > 0,
\ea
satisfies the  Hartle-Hawking boundary condition
\cite{HH}. The special cases
of this formula were considered in refs.
\cite{Zh1} (1-curvature case) and \cite{BIMZ}  ($\Lambda$-term case).

{}From the equation (4.3) it follows  that in the case
$A < 0$ a Lorentzian region exists as well as an Euclidean one
for ${\cal E} > 0$. In the case $A > 0$ only the Lorentzian region
occurs for ${\cal E} \ge 0$ but for ${\cal E} < 0$ both regions
exist.
The wave functions (4.2), (4.4) with $A > 0$ and ${\cal E} < 0$ describe
transitions between the
Euclidean and Lorentzian regions, i.e. tunneling
universes.

\subsection{Quantum wormholes}

Here we consider only real values of $p_{i}$.
In this case we have  ${{\cal E}} \geq 0$.

If $A >0$ the wave function $\Psi$ (4.2) is not
exponentially damped when $v \rightarrow \infty$, i.e. the condition
(i) for quantum wormholes (see the Introduction) is not satisfied.
It oscillates and may be interpreted
as corresponding to the  classical Lorentzian solution.

For $ A < 0$, the wave function (4.2) is exponentially damped for
large  $v$ only, when $B=K$ in (4.4). (We recall that
$$ {I_{\nu}}(z) \sim \frac{e^{z}}{\sqrt{2\pi z}},  \qquad
{K_{\nu}}(z) \sim \sqrt{\frac{\pi}{2z}}e^{-z},   $$
for  $z \rightarrow \infty$).
But in this case the
function $\Phi$ oscillates an infinite number of times, when $v
\rightarrow 0$. So, the condition (ii) is not satisfied. The wave function
describes the transition between Lorentzian and Euclidean regions.

The functions
\be 
{\Psi_{\vec{p}}}(z)=
e^{i\vec{p}\vec{z}}{K_{i|\vec{p}|/q}}
(\sqrt{-2A} q^{-1} e^{qz^{0}}),
\ee
may be used for constructing quantum wormhole solutions.
Like in \cite{CGar,Gar,IM4,BIMZ}  we consider superpositions of
the singular  solutions
\be  
{\hat{\Psi}_{\lambda,\vec{n}}}(z)=\frac{1}{\pi}
\int_{-\infty}^ {+\infty} dk {\Psi_{qk \vec{n}}}(z)e^{-ik \lambda},
\end{equation}
where $\lambda \in R$ and $\vec{n} \in S^{n-1}$ is a unit vector
($\vec{n}^{2}=1$).
The calculation gives
\be 
{\hat{\Psi}_{\lambda,\vec{n}}}(z)
=exp[-\frac
{\sqrt{-2\Lambda}}{q}e^{qz^{0}} \cosh(\lambda-q\vec{z}\vec{n})].
\end{equation}
It is not difficult to verify that the formula (4.12) leads to
the solutions of the WDW equation (4.1), satisfying the quantum
wormholes boundary conditions.

We note that the functions
\be 
\Psi_{m,\vec{n}}={H_{m}}(x^{0}){H_{m}}(x^{1})\exp[-\frac
{(x^{0})^{2}+(x^{1})^{2}}{2}] ,
\end{equation}
where
\ba 
&&x^{0}=(2/q)^{1/2}(-2A)^{1/4} \exp(qz^{0}/2) \cosh(
\frac{1}{2}q \vec{z}\vec{n}),  \\
&&x^{1}=(2/q)^{1/2}(-2 A)^{1/4} \exp(qz^{0}/2) \sinh(
\frac{1}{2}q \vec{z}\vec{n}),
\end{eqnarray}
$m=0, 1, \ldots, $ are also solutions of the WDW equation
with the quantum wormhole boundary conditions. Solutions of such
type were previously considered in  \cite{HawP,Zh2,Zh3,IM4,BIMZ}.
They are called the discrete spectrum quantum wormholes (see
\cite{Gar}) (and may form a basis in
the Hilbert space of the system \cite{Mar}).

Thus, in the case considered  the quantum wormhole solutions
(with respect to quasi-volume (4.5))
exist for the matter with a negative density (3.2) ($ A < 0$).


\section{ Third quantized model.}

Here we put $A >0$, i.e. the density of matter is positive.
We consider the case of a real $\Psi$-field as in
\cite{Zh5} for simplicity.
The WDW equation (4.1) corresponds to the  action
\be 
S = \frac{1}{2} \int d^{n+1} z \Psi \hat H \Psi.
\ee
Let us consider two bases of the solutions of the WDW
equation
$\{ {\Psi_{in}}(\vec p), {\Psi^{*}_{in}}(\vec p) \}$,
$\{ {\Psi_{out}}(\vec p), {\Psi^{*}_{out}}(\vec p) \}$
\ba 
&&{\Psi}_{in}(\vec p) = {\Psi}_{in}(\vec p, z) =
\left[ \frac{\pi}{2q \sinh(\pi |\vec{p}|/q)} \right]^{1/2}
{J_{-i |\vec{p}| /q}} \left(\frac{\sqrt{2A}}{q} e^{qz^0} \right)
(2 \pi)^{-n/2} \exp(i \vec{p} \vec z) \\
&&{\Psi}_{out}(\vec p)=
{\Psi}_{out}(\vec p, z)=
\frac{1}{2} \left(\frac{\pi}{q} \right)^{1/2}
{H^{(2)}_{i |\vec{p}| /q}} \left(\frac{\sqrt{2A}}{q} e^{qz^0} \right)
(2 \pi)^{-n/2} \exp(i \vec{p} \vec z)
\ea
where $J_{\nu}$  and $H^{(2)}_{\nu}$  are the Bessel and Hankel functions
respectively. These solutions
are normalized by the following conditions
\begin{equation} 
\left({\Psi_{in}}(\vec{p}), {\Psi_{in}}(\vec{p}') \right) = \left(
{\Psi_{out}}(\vec{p}), {\Psi_{out}}(\vec{p}') \right) = \delta \left(\vec
p-{\vec p\,}'\right) \ee
where
\be 
\left  (\Psi _1, \Psi _2 \right)=
i \int d^n \vec{z} \left(\Psi_1^{*}
\stackrel{\leftrightarrow}{\partial_{0}} \Psi_2 \right)
\end{equation}
is the charge form (indefinite scalar product). Here $\Psi_1
\stackrel{\leftrightarrow}{\partial}\Psi_2= \Psi_1 \,\partial \Psi_2-
\left(\partial \Psi_1 \right) \Psi _2$.  Due to asymptotic
behaviour
\ba 
&&{\Psi_{in}}(\vec{p}, z) \sim  {c_{in}}(|\vec{p}|) \exp(i \vec{p} \vec z
- i |\vec{p}| z^0), \qquad v \rightarrow 0, \\
&&{\Psi_{out}}(\vec{p}, z) \sim  \frac{{c_{out}}(|\vec{p}|)}{\sqrt{v}}
\exp(i \vec{p} \vec z  - i \frac{\sqrt{2A}}{q} v), \qquad
v \rightarrow + \infty.
\ea
where  ${\Psi_{in}}(\vec{p},z)$ and ${\Psi_{out}}(\vec{p}, z)$  are
negative frequency modes of "Kasner"-  and "Milne"- types respectively.

The standard quantization procedure \cite{BirD,GrMM} give us
\ba  
 {\Psi}(z)
 &&= \int d^n  \vec{p}
 \left [ {a_{in}^{+}}(\vec{p}){\Psi_{in}^{*}}(\vec{p},z) +
 {a_{in}}(\vec{p}){\Psi_{in}}(\vec{p},z) \right ]  \nonumber \\
  &&= \int d^n  \vec{p}
 \left [ {a_{out}^{+}}(\vec{p}){\Psi_{out}^{*}}(\vec{p},z) +
 {a_{out}}(\vec{p}){\Psi_{out}}(\vec{p},z) \right ],
\ea
where the non-trivial commutators read
\be  
[{a_{in}}(\vec{p}), {a_{in}^{+}}(\vec{p'})] =
[{a_{out}}(\vec{p}), {a_{out}^{+}}(\vec{p'})] =
\delta \left(\vec p-{\vec p\,}'\right).
\ee
"In" and "out" vacuum states satisfy the relations
\be 
{a_{in}}(\vec{p}) |0, in> = {a_{out}}(\vec{p}) |0, out> =0.
\ee
The modes (5.2) and (5.3) are related by the Bogoljubov transformation
\begin{equation} 
{\Psi_{in}}(\vec p) =
{\alpha}(|\vec{p}|){\Psi_{out}}(\vec p)
+ {\beta}(|\vec{p}|) {\Psi_{out}^{*}}(\vec p)
\end{equation}
\begin{equation} 
{\alpha}(\vec{p}) =
\left [ \frac{\exp(\pi |\vec{p}| /q)}{2 \sinh (\pi |\vec{p}|/q )}
\right]^{1/2} ,
{\beta}(\vec{p}) =
\left [ \frac{\exp( - \pi |\vec{p}| /q)}
{2 \sinh (\pi |\vec{p}|/q )} \right]^{1/2}. \ee

The vacuums $|0, in>$ and $|0, out>$ are unitary non-equivalent.
The standard calculation  \cite{BirD,GrMM} gives for a number
density  of "out-universes" (of "Milne- type") containing in the
"in-vacuum" ("Kasner-type" vacuum)
\be  
{n}(\vec{p})= |{\beta}(\vec{p})|^2 =
\left( \exp ( 2 \pi |\vec{p}|/q ) - 1 \right )^{-1}.
\ee
So, we obtained the Planck distribution with the temperature
\be 
T_{Pl} =  q/ 2 \pi = \sqrt{- <u,u>_{*}}/4 \pi .
\ee
The temperature (5.14)  depends upon the vector
$u = (u_i)$ ( i.e. on the equation of state):  $T_{Pl} = {T_{Pl}}(u)$.
For example, we get ${T_{Pl}}(u^{(\Lambda)})  = 2{T_{Pl}}(u^{(dust)})$.
In the Zeldovich matter limit  $u \rightarrow  0$ we have
$T_{Pl} \rightarrow +0$.

Remark 5. In \cite{I2} a regularization of propagators (in quantum
field theory) was introduced using the complex signature matrix
\be  
({\eta_{ab}}(w))  = diag(w,1, \ldots,1),
\ee
where $w \in C \setminus (-\infty, 0]$ is the complex parameter (Wick
parameter). Originally path integrals are defined
(in covariant manner) for $w > 0$ (i.e. in Euclidean-like region)
and then should be analytically continued to negative $w$. The
Minkowsky  space limit corresponds to $w = -1 + i0$ (in notations
of \cite{I2} $w^{-1} = -a$). The prescription \cite{I2} is a natural
realization of the Wick's rotation. In \cite{I3}
the analogs of the Bogoljubov-Parasjuk
theorems \cite{Bog} for a wide class of propagators regularized by
the complex metric (5.15) were proved. This formalism may be applied
for third-quantized models of the multidimensional cosmology. In this
case the corresponding path integrals should be analytically
continued  from the interval $1 < D< 2$ ($D$ is dimension), where
minisuperspace metric (2.12) is Euclidean, to $D = D_0 - i0$,
$D_0 = 1 + \sum_{i=1}^{n} N_i$. We note also that recently
J.Greensite proposed an idea of considering the space-time
signature as a dynamical degree of freedom \cite{Greens}
(see also \cite{CGr,EOR}).

\section{Appendix}

{\bf Proof of Proposition 2.}
We introduce the  new "diagonalized" variables
\be 
\beta^a = e^a_i \beta^i, \qquad  u_a = e_a^i u_i, \qquad
v_a = e_a^i v_i
\ee
where  matrices $(e_{a}^{i})$,  $(e^{a}_{i})$   satisfy the relations
(3.11)-(3.12) and (3.14).  From (3.12)-(3.14)  we have
\be 
(u_a) = (2q, \vec{0})  \qquad (\sigma^a) = (\sigma^i e^a_i) =
(q^{-1}, \vec{0})
\ee
and consequently  (see (3.96))
\be 
0= \beta^{i} u_i = \beta^{a} u_a = 2q \beta^{0} \ \Rightarrow \
(\beta^{a}) = (0, \vec{\beta}).
\ee
{}From the second relation in (3.96) we get
\be 
G_{ij} \beta^{i} \beta^{j} = \eta_{ab} \beta^{a} \beta^{b}
= \vec{\beta}^2 \leq 1/q^2.
\ee
For the vector $(v_a) = (v_0, \vec{v})$
we have $- v_0^2 +  \vec{v}^2 = <v,v>_{*} < 0$ and hence
\be 
|v_0| > |\vec{v}|, \qquad v_0  \neq 0.
\ee
We also
obtain from (6.2) and (6.5)
\be 
<u,v>_{*}  = - u_0 v_0 = - 2 q v_0   \neq  0.
\ee
Using relations (6.2), (6.3) and (6.5) we get
\be 
(\sigma^i + \beta^i) v_i = (\sigma^a + \beta^a) v_a =
=  q^{-1} v_0 +   \vec{\beta} \vec{v}
=  q^{-1} v_0 (1 +  \frac{q}{v_0} \vec{\beta} \vec{v}).
\ee
Eqs. (6.4), (6.5) imply the following inequality
\be 
|\frac{q}{v_0} \vec{\beta} \vec{v}| \leq
\frac{|\vec{v}|}{v_0} q |\vec{\beta}| \leq
\frac{|\vec{v}|}{v_0}  < 1.
\ee
{}From (6.6)-(6.8) (and $q > 0$) we get the proposed identity  (3.97).
The proposition is proved.

\pagebreak

\pagebreak

Fig. 1. The graphical representation of the allowed cone
       $<u,u>_{*} < 0$ and different domains in it
         on the plain $\xi_i = p_i/\rho$, $i =1,2$,
        for the case $n =2$ and $N_1 = 3$, $N_2 = 6$.
       The hyperbola $\bar{\sigma} =1$ corresponds to
       the exponential inflation (3.86).

\end{document}